\documentclass[12pt,a4paper]{article}
\usepackage{graphicx,fancyhdr}
\usepackage{cite,citesort}
\usepackage{color}
\usepackage{wasysym}
\begin{document}

\pagenumbering{arabic}

\vspace{-40mm}

\begin{center}
\Large\textbf{Direct writing of CoFe alloy nanostructures by focused electron beam induced deposition from a heteronuclear precursor}\normalsize \vspace{5mm}

\large F. Porrati${^1}$, M. Pohlit${^1}$, J. M\"{u}ller${^1}$, S. Barth${^2}$, F. Biegger${^2}$, C. Gspan${^3}$, H. Plank${^3}$, and M. Huth${^1}$

\vspace{5mm}

\small\textit{1. Physikalisches Institut, Goethe-Universit\"at, Max-von-Laue-Str.~1, D-60438 Frankfurt am Main, Germany}

\small\textit{2. Vienna University of Technology, Institute of Materials Chemistry, Getreidemarkt 9/BC/02, A-1060 Wien, Austria}

\small\textit{3. Institute for Electron Microscopy and Nanoanalysis, Graz University of Technology, A-8010 Graz, Austria}

\newpage

\end{center}

\vspace{0cm}
\begin{center}
\large\textbf{Abstract}\normalsize
\end{center}

Recently, focused electron beam induced deposition has been employed to prepare functional magnetic nanostructures with potential in nanomagnetic logic and sensing applications by using homonuclear precursor gases like Fe(CO)$_5$ or Co$_2$(CO)$_8$. Here we show that an extension towards the fabrication of bi-metallic compounds is possible by using a single-source heteronuclear precursor gas. We have grown CoFe alloy magnetic nanostructures from the HFeCo$_3$(CO)$_{12}$ metal carbonyl precursor. The compositional analysis indicate that the samples contain about 80~at$\%$ of metal and 10~at$\%$ of carbon and oxygen. Four-probe magnetotransport measurements are carried out on nanowires of various sizes down to a width of 50~nm, for which the room temperature resistivity of 43~$\mu\Omega$cm is found. Micro-Hall magnetometry reveals that 50~nm$\times$250~nm nanobars of the material are ferromagnetic up to the highest measured temperature of 250~K. Finally, the TEM microstructural investigation shows that the deposits consist of a bcc Co-Fe phase mixed with a FeCo${_2}$O${_4}$ spinel oxide phase with nanograins of about 5~nm diameter.

\newpage

\begin{center}
\large\textbf{1. Introduction}\normalsize
\end{center}

In the last years, the synthesis of magnetic compounds at nanoscale has generated a large scientific interest due to potential applications in spintronics, thermoelectrics, nanoelectronics and information technology~\cite{schmitt10,wang07}. Among the various nanofabrication techniques, focused electron beam induced deposition (FEBID) is increasingly used in research and prototyping applications due to its direct writing and high resolution capabilities~\cite{utke08,huth12}. In FEBID the molecules of a precursor gas injected in a scanning electron microscope (SEM) decompose by interaction with the electron beam, forming a sample during the rastering process. Currently, FEBID relies on homonuclear precursors to fabricate nanostructures. Among the possibly relevant organometallic precursors, most attractive are those which result in virtually complete dissociation under electron impact. In this case, full desorption of the organic ligands under suitable FEBID process parameters can in principle be realized resulting in clean metallic deposits. Additional design goals for the ideal FEBID precursor are sufficient thermal stability and suitable vapor pressure, i.e., in the range of 10$^{-2}$ to 50~mbar, at about room temperature. The precursors Fe(CO)$_5$ and Co$_2$(CO)$_8$ have shown to yield high metal content deposits, and functional magnetic iron and cobalt nanostructures with lateral size below 30~nm have been grown~\cite{deteresa14,serrano11,gavagnin13}. However, great care has to be taken in utilizing these precursors, since both are inclined to spontaneous dissociation on activated surfaces, as has been shown in several studies~\cite{walz,muthu}. This indicates the complexity of the task in finding the delicate balance between high metal yield precursors which, at the same time, must be sufficiently stable against spontaneous dissociation. As a second limitation in FEBID, in spite of the many successes in fabrication, characterization and application of nanostructures grown by this technique in various areas~\cite{huth12,utke2002,gabureac2010,serrano11}, the number of available magnetic materials is still very limited. Mixing precursor gases for the fabrication of binary alloys was proposed about a decade ago~\cite{che05} and more recently realized by employing dual- or multichannel precursor gas injection systems~\cite{winhold11,fab12,fab13}. However, the exact and reproducible control of the deposits' composition remains a challenge and, more importantly, in the few examples studied so far the overall metal content in the deposits has been found to be below about 60$\%$~\cite{fab13}.

Here we show that magnetic alloy nanostructures can be fabricated reproducibly by FEBID with well-defined elemental composition and high metal content by using a heteronuclear precursor. We use the carbonyl HFeCo$_3$(CO)$_{12}$ to fabricate FeCo alloy phase nanostructures with metal content up to 84$\%$  and lateral size down to 50~nm. Magnetotransport measurements and Hall magnetometry reveal the ferromagnetic nature of the deposit up to the highest measured temperature of 250~K. By transmission electron microscopy (TEM) we identify the metallic phase, verify the Co$_3$Fe elemental composition and characterize a residual oxide phase.

\begin{center}
\large\textbf{2. Experimental section}\normalsize
\end{center}

\textbf{2.1 Precursor.} \textit{Synthesis of HFeCo$_3$(CO)$_{12}$}. The synthesis was carried out in a slightly modified procedure as compared to that described by Chini et al.~\cite{chini}. The synthesis and handling of the precursor requires Schlenk techniques  to prevent oxidation. The solvents were degassed before use. All chemicals were purchased from Sigma Aldrich.

3.01 g (8.8 mmol) Co$_2$(CO)$_8$ were mixed with 1.02 g (5.2  mmol) Fe(CO)$_5$  in a round bottom flask. Subsequently, 9.5 ml acetone were added drop wise and the mixture was stirred at 40$^\circ$~C for 2~h and 12~h at 60$^\circ$~C. Afterwards, the volatile components were removed under reduced pressure (10$^{-2}$~mbar; 25$^\circ$~C) from the initially dark reddish-brown solution and collected in a cooling trap (liquid N$_2$). In the next step, 20~ml purified water were added to the dark solid and the liquid phase was filtered in a flask containing  30~ml HCl (37~$\%$). The yield was increased by adding 20~ml water to the remaining solid and filtering this solution in the same HCl containing flask. A dark purple solid was formed immediately in the acidic solution, which was stirred for 2~h before the liquid was removed by filtration. The solid was dried in a desiccator over P$_2$O$_5$ for 12~h and recrystallized from toluene to obtain highly crystalline HFeCo$_3$(CO)$_{12}$.

\textit{Characterization}. The HFeCo$_3$(CO)$_{12}$ was analyzed by IR, powder XRD and single crystal diffraction, which was used to provide a reference for the powder XRD characterization. The CIF file for the results of the single crystal structure determination is not included or deposited in a database due to the uncertainty of assigning Fe and Co to specific positions (available from the authors upon request). However, the three equivalent positions should be occupied by Co atoms as shown in figure S1 of Supplementary Data. Strong disorder is observed for the crystals with tetrahedra tilted by 180$^\circ$ and 50$\%$ population density of the specific positions. The simulated powder XRD (Diamond 3) using the single crystal data is shown as a reference for the experimentally obtained pattern (figure S2) using Cu K$\alpha$. Texture effects due to large anisotropic crystallites have not been considered. The strongly anisotropic growth of crystals (needle morphology) is the reason for the weak reflexes at 13 and 14 degrees. However, the positions of the most prominent reflexes are the same. IR spectra do not show any CH nor OH related signals.  ATR measurements show strong absorption bands at 2004, 1966, 1863 and 1104 cm$^{-1}$ for this compound.

\textbf{2.2 Fabrication.} The samples were grown by using a dual beam SEM/FIB microscope (FEI, Nova NanoLab 600), equipped with a Schottky electron emitter. The HFeCo$_3$(CO)$_{12}$ precursor was heated at 64$^{\circ}$ for half an hour before use. The basis pressure of the SEM was 4.1$\cdot$10$^{-6}$ mbar, which increased to 4.2$\cdot$10$^{-6}$ mbar during deposition. The precursor was injected in the SEM via a capillary with 0.5~mm inner diameter in close proximity to the focus of the electron beam on the surface substrate. The distance capillary-surface substrate was about 100~$\mu m$. The samples were grown on Si(p-doped)/SiO2(10 nm)/Si3N4(100 nm) substrates and contacted with Au(60 nm)/Cr(20 nm) electrodes prepared by UV photolithography. CoFe alloy nanowires were grown with two different electron beam parameters. Nanowires with ca. 700~nm width, 90~nm height and 5~$\mu$m length, were prepared with a beam energy, beam current, pitch, and dwell time of 5~keV, 2.9~nA, 20~nm, and 1~$\mu$s, respectively. The growth rate of the nanowires was about 18~nm/min. The good growth rate is compatible with a high electron induced dissociation cross section and a small sticking coefficient. Small nanowires with ca. 50~nm width, 10~nm height and 300~nm length, were prepared with a beam energy, beam current, pitch, and dwell time of 5~keV, 30~pA, 10~nm, and 100~$\mu$s in the high resolution deposition mode, respectively. The latter parameters were used also to deposit the nanobars with 50~nm width, 250~nm length and 30~nm height for Hall magnetometry measurements. AFM measurements were carried out in the non-contact mode (Nanosurf, easyscan2). The composition of the deposits was given by energy dispersive x-ray analysis (EDX).

\textbf{2.3 Electrical and magnetotransport measurements.} Four-point electrical transport measurements were carried out in the temperature range 2-300~K in a variable-temperature insert mounted in a $^4$He cryostat equipped with a 12~T superconducting solenoid. Standard measurements were performed using a Keithley Sourcemeter 2400 and an Agilent 34420A nanovoltmeter. For measuring the smallest nanowire, we used a four-wire set-up employing a SR830 lock-in amplifier, a SR560 pre-amplifier and an Eaton RT-20A tunable voltage divider.

\textbf{2.4 Hall magnetometry.} Six adjacent Hall-crosses with areas of 1$\times$1~$\mu m^2$ were defined by electron beam lithography followed by wet chemical etching. (The electronically active area is slightly reduced due to edge depletion). The micro-Hall sensor was fabricated from a two dimensional electron system (2DES) based on an AlGaAs/GaAs heterostructure which lies about 140~nm below the wafer surface~\cite{wirth99,pohlit2015}. The sensitive 2DES of the Hall structure is covered by a thin top gate (5~nm Cr/40~nm Au) and electronically contacted by annealed planar AuGe/Ni contacts. The CoFe sample was deposited slightly off-center on one cross to maximize the generated Hall voltage V$_H$ = I/ne $\cdot$ $<B_z>$ during magnetic reversal (see Fig.~S3), where n and I denote carrier density and applied current, respectively, and $<B_z>$ the magnetic stray field emanating from the end of the sample averaged over the active area of the Hall cross. After deposition the sensor was transferred to a low temperature cryogenic system equipped with a superconducting solenoid. Above 75~K a variable temperature insert (VTI) cryostat was used. The stray field of the sample, which in first approximation is directly proportional to its magnetization, is detected by measuring the Hall voltage generated in the sensor plane using a standard low-frequency Lock-In technique. The sample's magnetization is switched by an external magnetic field parallel to the sensor plane, which ideally has no perpendicular component and therefore creates no Hall signal. A small misalignment ($\sim~1^\circ$) can be corrected for, and the sensitivity can be further improved by eliminating any background signal by subtracting an empty reference cross (the sensor structure consists of an array of six identical Hall crosses).

\textbf{2.5 TEM.} The TEM-lamellae were prepared with a NOVA 200 Nanolab dual beam system from FEI. TEM investigations were carried out on a Tecnai F20 from FEI with a Schottky Field Emitter at 200 kV. Images were performed with a post column energy filter (Gatan Imaging Filter, GIF) using an energy slit of 10 eV. The images were recorded zero-loss filtered (i.e. elastically scattered electrons only) on a 2k CCD. For the image recording and processing (Fourier transformation) the software DigitalMicrograph from Gatan was used.

\begin{center}
\large\textbf{3. Results}\normalsize
\end{center}

\begin{figure}\center{\includegraphics[width=12cm]{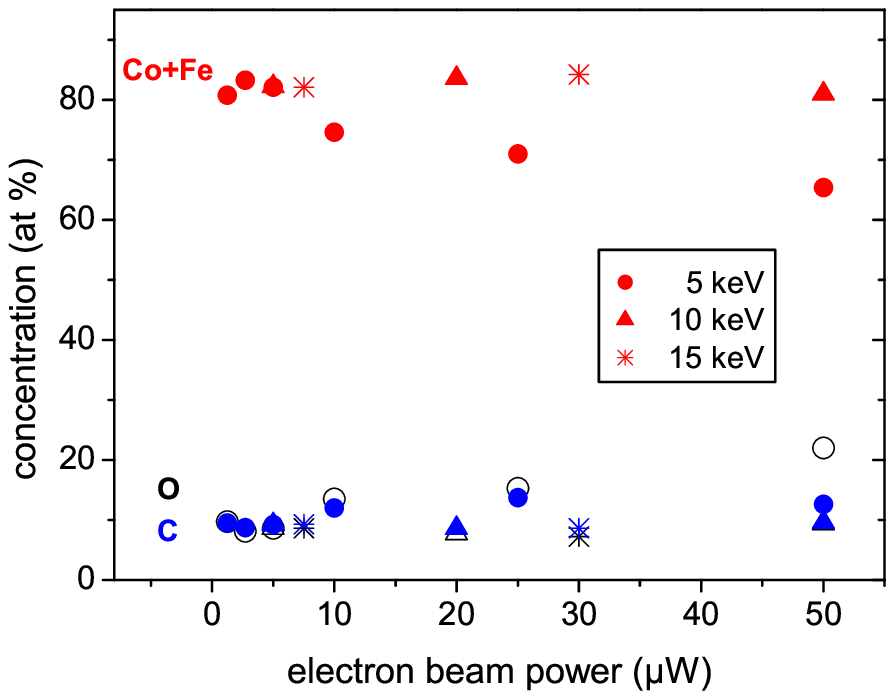}}
\caption{Atomic concentrations vs electron beam power used during deposition. EDX measurements were carried out with 5~keV electron beam acceleration voltage, in order to avoid the contribution of photons from the substrate. Since the L-lines of Co and Fe partially overlap, the independent evaluation of [Co] and [Fe] is unprecise. Therefore the sum [Co]+[Fe] is plotted. The precursor stoichiometric value [Co]/[Fe]=3 is found by considering the K-lines, as proved in the TEM investigation (see Figure S8).}\label{concentration2}\end{figure}

\textbf{3.1 Deposit composition.} The material was characterized by energy dispersive X-ray analysis (EDX), carried out on deposits of 2$\times$2~$\mu$m$^2$ lateral size and thickness in the range of 100~nm. In Fig.~\ref{concentration2} we plot the concentrations of the elements as function of the electron beam power. (Fig.~S4 in the Supporting Data depicts the concentrations versus the electron beam current). In the range of the electron beam parameters considered, the global metal content of the deposits, i.e., [Co]+[Fe], is typically about 80~at$\%$, with a peak value of 84~at$\%$. Fig.~S5 depicts a typical EDX spectrum with about 80~at$\%$ [Co]+[Fe] and 10~at$\%$ [O] and [C].

After having performed the compositional analysis on the square deposits, we fabricated two kinds of functional magnetic nanostructures: magnetic nanowires, which were characterized by magnetotransport measurements and transmission electron microscopy, and nanobars, which were used to characterize the material by micro-Hall magnetometry.

\begin{figure}\center{\includegraphics[width=12cm]{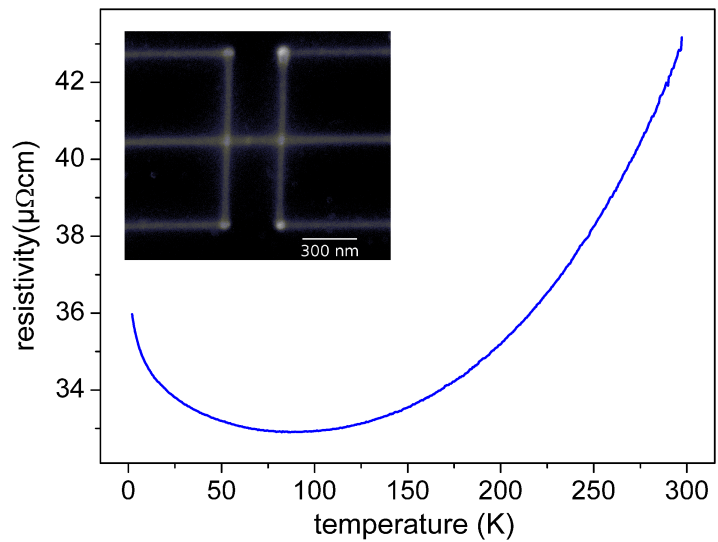}}
\caption{Temperature dependent resistivity of a nanowire with 50~nm width and 10~nm thickness (sample A), as measured by AFM. Inset: SEM image of an exemplary nanowire written with the same electron parameters used to prepare the nanowire measured in transport.}\label{nanowire_A}\end{figure}

\textbf{3.2 Electrical properties.} Transport measurements were carried out on a set of three nanowires. The smallest one, sample A, see inset of Fig.~\ref{nanowire_A}, has a width of 50~nm and a thickness of 10~nm. The electrical resistivity at room temperature is about 43~$\mu\Omega$cm. This value is close to the one obtained for cobalt nanowires fabricated by FEBID~\cite{fernandez09}, about a factor 2 smaller than the value of iron FEBID structures~\cite{lavrijsen11,fab11} and about a factor 7 and 4 larger than the values of bulk cobalt and iron\cite{lavrijsen11,fab11}, respectively.
The higher resistivity value of the nanowire compared to those of bulk samples is expected to be due to the residual impurities of C and O and to the surface scattering, as, e.g., described in the theory of Fuchs and Sondheimer~\cite{sondheimer}. The resistivity of sample A grows monotonically in the range 100~K to 300~K as expected for metallic samples, see Fig.~\ref{nanowire_A}, and shows a minimum at 87~K. A similar behavior is known from thin epitaxial Fe films indicating a transition from 3D to 2D behavior at lower temperature, and has been attributed to weak electron-localization and/or electron-interaction effects~\cite{rubinstein,sangiao}. The residual resistivity ratio (RRR) between room temperature and 2~K is about 1.2, slightly smaller than those obtained for highest metal content~\cite{fernandez09} or post growth purified~\cite{begun15} FEBID cobalt nanowires. The resistivity temperature dependence of the larger nanowires, samples B and C, is reported in Fig.~S6 of the Supporting Information.

\begin{figure}\center{\includegraphics[width=14cm]{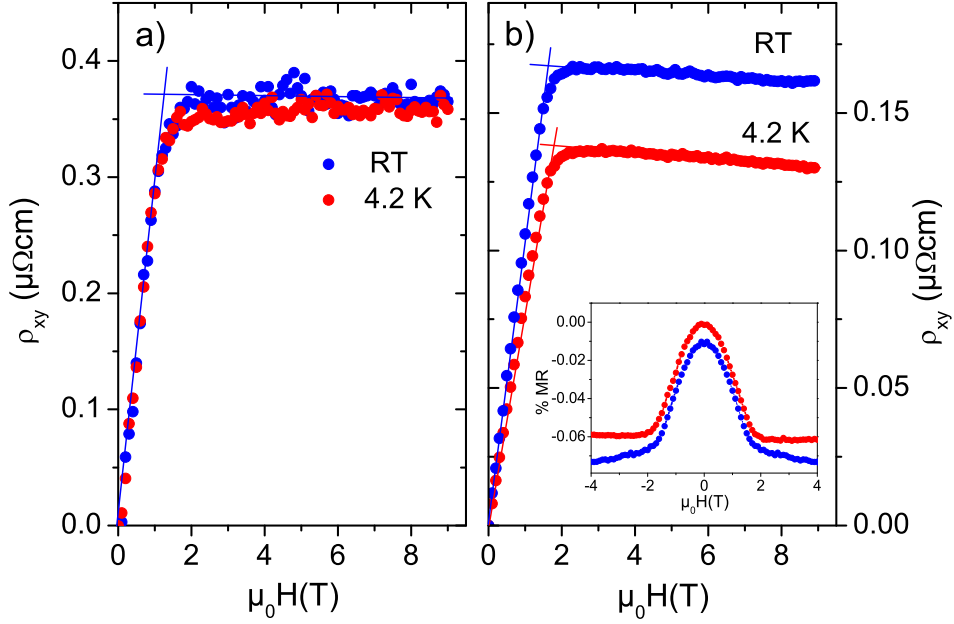}}
\caption{Hall resistivity $\rho_{xy}$ for samples A (panel a) and C (panel b). The intersection of the slopes of the Hall resistivity at low and high temperature gives the saturation field $M_s$, which is 1.2~T and 1.6~T for sample A and C, respectively. Inset: perpendicular magnetoresistance of sample C. The rotation of the magnetic moments of the sample towards the direction of the applied field increases the electron scattering probability leading to the negative magnetoresistance. In the figures, red and blue lines are the measurements at 4.2~K and RT, respectively.}\label{Hall_effect}\end{figure}

\textbf{3.3 Magnetotransport properties.} In Fig.~\ref{Hall_effect} we plot the Hall resistivity of samples A and C. The Hall resistivity is given by the sum of the ordinary and of the anomalous Hall effects, $\rho_{xy}$=$\rho_{OR}$+$\rho_{AN}$, with $\rho_{OR}$=$R_0\mu_0H$ and $\rho_{AN}$=$R_s\mu_0M$, being $\mu_0H$ the magnetic induction corresponding to the external field $H$, $R_0$ and $R_s$ the ordinary and the anomalous Hall coefficients, respectively, and $M$ the magnetization of the sample in the field direction assuming a demagnetizing factor of the nanowires of $N\approx1$~\cite{hubert}. The ordinary Hall coefficient $R_0$ is determined by the slope of the Hall effect at high magnetic fields, while $R_s$ is given by the extrapolation of this slope to zero field. In general, according to a unified theory for multiband ferromagnetic metals with dilute impurities~\cite{onoda}, the anomalous Hall effect (AHE) shows a crossover from extrinsic to intrinsic behavior by lowering the longitudinal conductance $\sigma_{xx}$. In particular, in the limit of highly conductive metals, i.e., for $\sigma_{xx}>10^6~\Omega^{-1} cm^{-1}$, the AHE is dominated by the extrinsic skew scattering. Here $\sigma_{xy}\propto\sigma_{xx}$, being $\sigma_{xy}$ the anomalous conductivity. In the intermediate metallic region, with $\sigma_{xx}=10^4-10^6~\Omega^{-1} cm^{-1}$, the anomalous conductivity is constant, i.e., $\sigma_{xy}$=const., and it is dominated by the contribution of the intrinsic Berry-phase~\cite{onoda,nagaosa}. Finally, for $\sigma_{xx}<10^4~\Omega^{-1} cm^{-1}$, in the dirty metal region, $\sigma_{xy}\propto\sigma_{xx}^{1.6}$. The anomalous and the longitudinal conductivities of our samples are extracted from Fig.~\ref{Hall_effect}, taking into account that $\sigma_{xy}\simeq\rho_{xy}/\rho_{xx}^2$ and $\sigma_{xx}\simeq1/\rho_{xx}$, for $\rho_{xy}\ll\rho_{xx}$. For sample A we find $\sigma_{xx}=2.3\cdot10^4~\Omega^{-1} cm^{-1}$ and $\sigma_{xy}=1.9\cdot10^2~\Omega^{-1} cm^{-1}$. These values, which show that sample A is in the intermediate metallic regime, are in the same range as those measured for Fe-based and Co-based FEBID nanowires~\cite{fab11,fernandez09,begun15}. For sample C we find $\sigma_{xx}=0.85\cdot10^4~\Omega^{-1} cm^{-1}$ and $\sigma_{xy}=12~\Omega^{-1} cm^{-1}$. Similar values are found for sample B. Therefore the transport properties of sample B and C are those of a dirty metal, similarly to non-purified Co-based FEBID~\cite{begun15} nanostructures and to Fe epitaxial thin films~\cite{sangiao}. Finally, we notice that the value of the anomalous Hall resistivity of sample A is about a factor 3 larger than the value of sample C, see Fig.~\ref{Hall_effect}. This increase, which is also found in Co-based FEBID deposits~\cite{fernandez09,serrano11}, can be attributed to electron surface scattering due to the reduced size of the samples~\cite{gerber2002}.

Magnetoresistance measurements were carried out on samples B and C in the perpendicular geometry, i. e., with the magnetic field perpendicular to the sample surface, as for the Hall effect measurements. The low signal to noise ratio hindered measurements on sample A. Negative magnetoresistance with values smaller than one $\permil$ were found, see the inset of Fig.~\ref{Hall_effect}. These values are about one order of magnitude smaller than those found in Co-based FEBID nanowires~\cite{fernandez09}, which we attribute to the higher resistivity of our samples. For the same reason the magnetoresistance of sample B is about 30$\%$ smaller than the one of sample C.

\textbf{3.4 Hall magnetometry.}

\begin{figure}\center{\includegraphics[width=12cm]{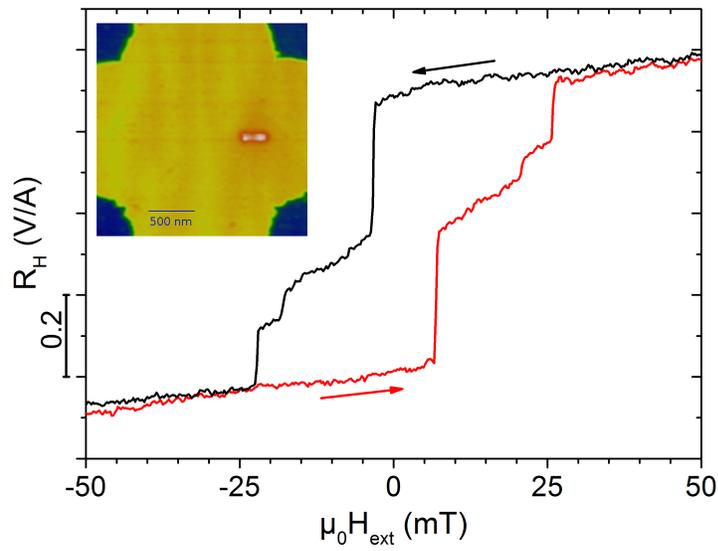}}
\caption{Magnetic hysteresis loop at T=15~K of the CoFe alloy nanobar showing a characteristic wasp-waist shape and distinct steps. An empty reference cross has been subtracted as background. Inset: AFM image of the Hall cross and the measured CoFe nanobar.}\label{Hall_magnetometry}\end{figure}

The magnetic properties of a single CoFe alloy nanobar element with 50~nm$\times$250~nm planar size and 30~nm height have been determined by micro-Hall magnetometry. At low temperatures a distinct hysteresis loop is observed during magnetization reversal proving the ferromagnetic character of the sample, see Fig.\ref{Hall_magnetometry}. The hysteresis has a characteristic wasp-waist shape with distinct step-like switching. While different mechanisms are known to cause such a characteristic constricted hysteresis shape, two fundamentally different explanations appear to be likely in the present case: the reversal via several metastable magnetic states involving the formation and the annihilation of magnetic vortices or the reversal of a mixed system consisting of a soft and a hard magnetic phase~\cite{hengstmann01,tauxe1996}. In the measured temperature range from 250~K down to about 300~mK the shape of the hysteresis remains essentially the same (see Fig.~S7). As a general trend, in particular below 15~K, the coercivity increases with decreasing temperature.

\textbf{3.5 Microstructural characterization.}

\begin{figure}\center{\includegraphics[width=12cm]{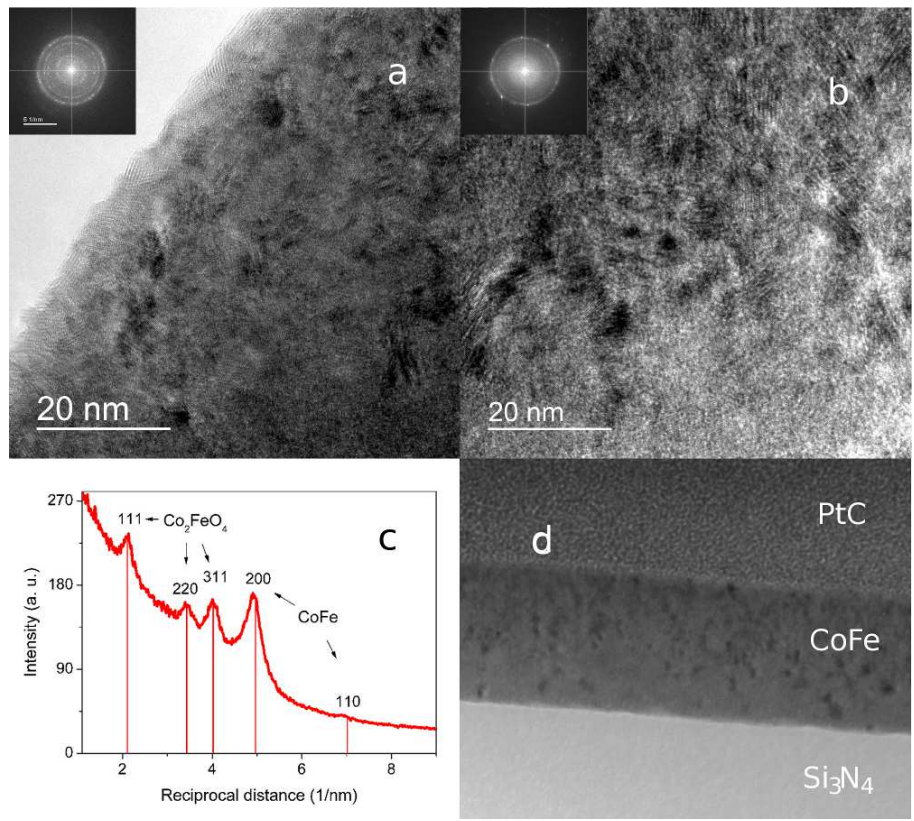}}
\caption{HRTEM micrographs of samples B (panel a) and C (panel b). In the inset the corresponding FFT are plotted. c) Radial intensity of the FFT of sample B. The peaks belong to the Co$_2$FeO$_4$ spinel oxide and to the bcc CoFe phases. d) Zoom out of the micrograph of sample C. The Pt-C cover layer, the CoFe sample and the Si$_3$N$_4$ substrate are visible, see text for details.}\label{TEM}\end{figure}

The microstructure of the alloy deposits was investigated by means of high-resolution TEM measurements on samples B and C. TEM lamellas were prepared after covering the samples with a Pt-C protecting layer by FEBID. In Fig.~\ref{TEM}a) we show a cross-sectional TEM micrograph of sample B. Dark metal nanocrystals with typical diameters of about 5~nm are homogeneously distributed in the sample. The corresponding Fast Fourier Transformation (FFT), see inset, shows well defined rings. In Fig.~\ref{TEM}c we depict the radial intensity obtained by azimuthal integration of the FFT. Two groups of peaks are visible: one compatible with the (111), (220) and (311) reflections of the spinel Co$_2$FeO$_4$, the other with the (220) and (110) reflections of CoFe. Note that the CoFe peaks are close to the (002) and (102) peaks of hcp-Co. However, the (100) and (101) hcp-Co peaks are absent, see the positions 4.58 1/nm and 6.72 1/nm in Fig.~\ref{TEM}c. Therefore, although we cannot exclude the presence of hcp-Co, the systematic absence of the (100) and (101) reflections indicates clearly that the sample contains only Co$_2$FeO$_4$ and CoFe nanocrystals. Differently from sample B, which after deposition was irradiated with an electron dose of about 0.12~$\mu C/\mu m^2 $ to decrease the electrical resistivity, sample C was irradiated only during the deposition of the additional Pt-C protecting layer. Therefore, while the entire thickness of sample B was subjected to the electron irradiation, only the upper part of sample C was irradiated during the growth of the protecting layer. In Fig.~\ref{TEM}b the TEM micrograph of sample C is shown. The nanocrystals are localized in the upper part of the picture, close to the Pt-C covering layer (see Fig~\ref{TEM}d), while the lower part shows an amorphous region of the sample. As a consequence, the corresponding FFT image does not have well-defined rings, see inset.

\newpage

\begin{center}
\large\textbf{4. Discussion}\normalsize
\end{center}

In general, nanocomposite materials prepared by FEBID are made of metallic nano-grains embedded in a carbonaceous matrix. Depending on the metal content of the nanocomposite, different electrical transport regimes are found and a variety of applications have been proposed~\cite{huth12,deteresa14}. Prototype devices for spintronics and nanoelectronics require high metal content materials, which, up to date, have been obtained only by means of a few homonuclear carbonyl precursors like Co$_2$(CO)$_8$, Fe(CO)$_5$ and Fe$_2$(CO)$_9$ ~\cite{utke2004,lukasczyk08,fernandez09,lavrijsen11,gavagnin13}. Recently, in order to expand the number of materials available, the fabrication of binary alloys obtained by mixing two precursor gases, has been considered ~\cite{che05}. However, this method is scarcely employed since it implies the use of non-standard multi-channel gas injection systems and a high-degree of control of the precursors' gas fluxes~\cite{winhold11,fab12}. Moreover, a maximum metal content below 60$\%$ was found and an electrical resistivity of about 10$^4$ $\mu\Omega cm$ was measured~\cite{fab13}, which is two orders of magnitude smaller than what is desirable for nanotechnology applications. A straightforward alternative is the one suggested here, which consists of the use of a single-source heteronuclear precursor gas. This route, known from the preparation of thin film alloys in chemical vapor deposition~\cite{boyd97}, allows the direct growth of magnetic nanostructures with the same stoichiometry as the heteronuclear precursor. By using the HFeCo$_3$(CO)$_{12}$ carbonyl we grow magnetic nanostructures with a metal content above 80~at$\%$, in a wide range of the electron beam power used. It is remarkable that these results are obtained in non-optimal vacuum conditions, i.e., with a base pressure of $4\cdot10^{-6}$~mbar. For comparison, we note that Co- and Fe-based deposits grown in non-optimal vacuum conditions have a metal atomic concentration about 15-20~$\%$ smaller than those found in deposits grown in optimal vacuum conditions, i.e., with a base pressure between 1 and $2\cdot10^{-6}$~mbar~\cite{lavrijsen11,fernandez09,begun15}.

We now turn to the microstructural investigations. After deposition the samples were covered with a Pt-C FEBID protecting layer in order to prevent oxidation and to avoid damages during the TEM lamella preparation. In sample C, the covering process induces a partial crystallization, mainly located in the upper layers, where the electron irradiation is more effective, see Fig.~\ref{TEM}d. The amorphous to crystalline transformation is more evident in sample B, which was specifically irradiated after deposition to increase the conductivity. From the TEM investigation of sample B, see Fig.~\ref{TEM}a, and, although less clearly, of sample C, see Fig.~\ref{TEM}b, it is evident that the material investigated contains Co-Fe and Co$_2$FeO$_4$ nanocrystals. Furthermore, from the TEM and the EDX analysis, we find that the ratio [Co]/[Fe] is equal to three and that the elements' concentrations are about 60~at$\%$ for Co, 20~at$\%$ for Fe, 10~at$\%$ for O and 10~at$\%$ for C. By making the assumption that Co$_2$FeO$_4$ binds all the oxygen atoms available, it follows that 7.5~at$\%$ of the metal present in the material belongs to the spinel oxide phase, while the remaining 72.5~at$\%$ forms the bcc Co-Fe crystal phase. Therefore one concludes that the Co-Fe nanocrystals phase present in the material is the dominant one.

In order to interpret the micro-Hall magnetometry data and in particular the characteristic wasp-waist shape of the hysteresis loop measured at low magnetic fields, see Fig.~\ref{Hall_magnetometry}, different scenarios should be considered. First, the reversal may take place via metastable states involving the formation and annihilation of magnetic vortexes, as found in Co and Fe circular disk-like structures~\cite{hengstmann01}. In particular, it is known that switching events between distinct micromagnetic configurations strongly affect the stray field emanating from the sample, which results in step-like changes in the hysteresis loop measured by micro-Hall magnetometry~\cite{hengstmann01}. Second, the behavior of the system during reversal may relate to the intrinsic nature of the material, i.e., to the presence of the CoFe$_2$O$_4$ and Co-Fe phases. Indeed, it is known from the literature that two-phase magnetic systems can generate a constriction in the hysteresis loop as found in our experiment~\cite{tauxe1996}. Note also that spinel oxide CoFe$_2$O$_4$ and Co$_2$FeO$_4$ thin films show such a wasp-waisted hysteresis too, which has been attributed either to a characteristic cusped surface morphology or to the presence of antiphase boundary defects~\cite{coll14}. However, according to the previous considerations, the Co$_2$FeO$_4$ spinel oxide phase is not the dominant one within our deposits, therefore such an explanation appears rather unlikely. Finally, it cannot be excluded that the reversal mechanism in our case is a complex process in which both the mixed phase nature of the material and the existence of multiple metastable states related to the constrained dimension of the nanostructure, play a major role.

\begin{center}
\large\textbf{5 Conclusions}\normalsize
\end{center}

In this work, we have fabricated CoFe alloy magnetic nanostructures by means of FEBID from the HFeCo$_3$(CO)$_{12}$ heteronuclear carbonyl precursor. A complete characterization of the composition, the microstructure, the electrical and magnetic properties of the deposits has been carried out. The bi-metallic and ferromagnetic nature of the nanocomposite, the high metal content obtained, together with the high resolution and direct-writing capabilities offered by the FEBID technique, makes these functional magnetic nanostructures attractive for prototype applications in spintronics and information technology. Furthermore, our results indicate that heteronuclear carbonyl precursors with multiple-metal bonding might form a very promising precursor class for future development of FEBID towards a direct-writing technique for multi-component metallic nanostructures.

\begin{center}
\large\textbf{6 Acknowledgments}\normalsize
\end{center}

Authors thank Roland Sachser for helping in the transport measurements of the 40~nm wide nanowire. FP acknowledge the financial support of the Deutsche Forschungsgemeinschaft under the project PO 1415/2-1. This work was carried out in the frame of the CELINA COST Action CM1301.

\newpage

\newpage

\setcounter{figure}{0}

\renewcommand{\thefigure}{S\arabic{figure}}

\vspace{30mm}

\begin{figure}[ht]\center{\includegraphics[width=10cm]{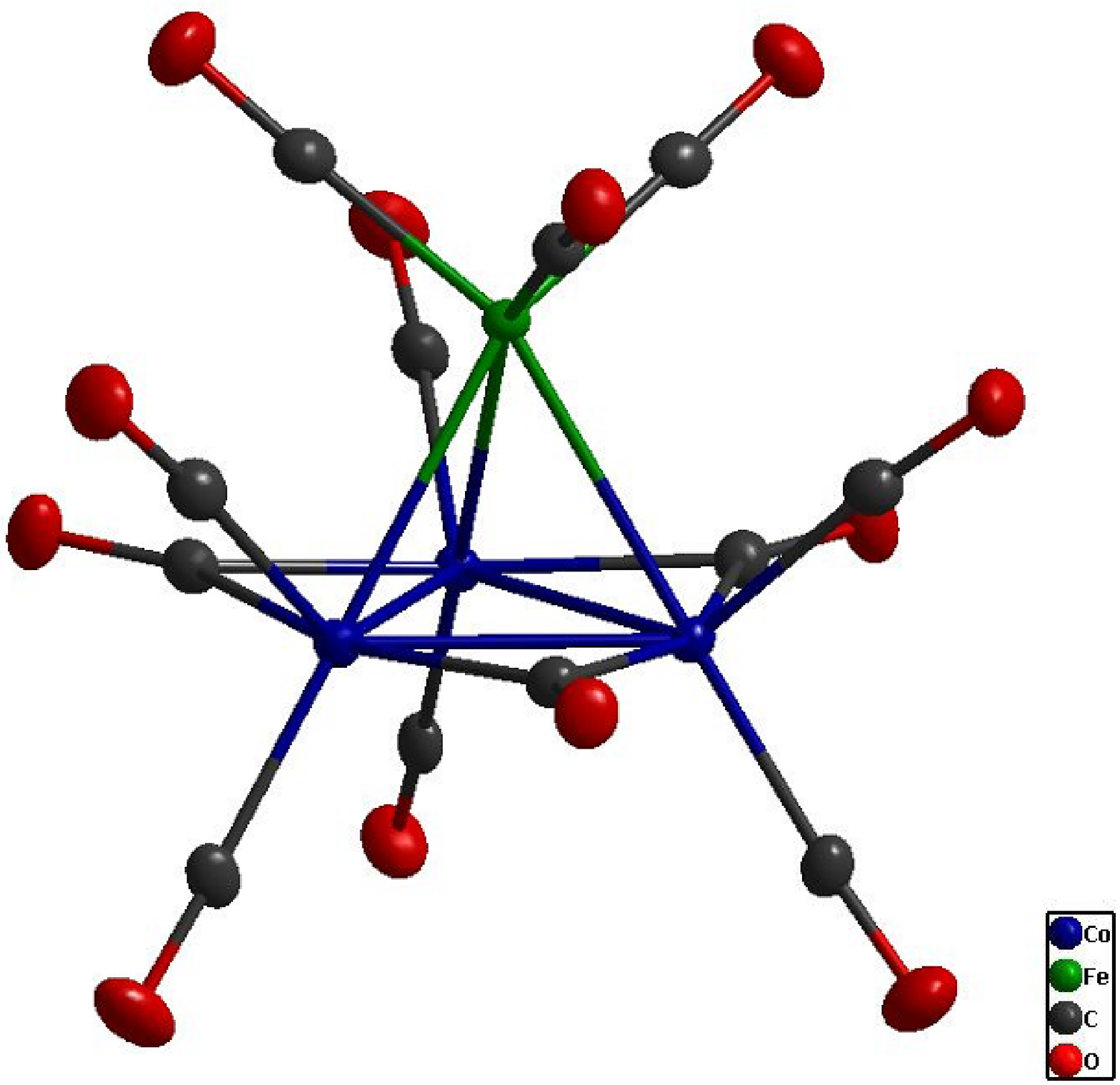}}
\caption{Structure of a HFeCo$_3$(CO)$_{12}$ molecule. A tetrahedral is made of one Fe atom and three Co atoms, respectively. Each atom bounds to three carbonyl groups.}
\label{Precursor structure}\end{figure}

\begin{figure}\center{\includegraphics[width=12cm]{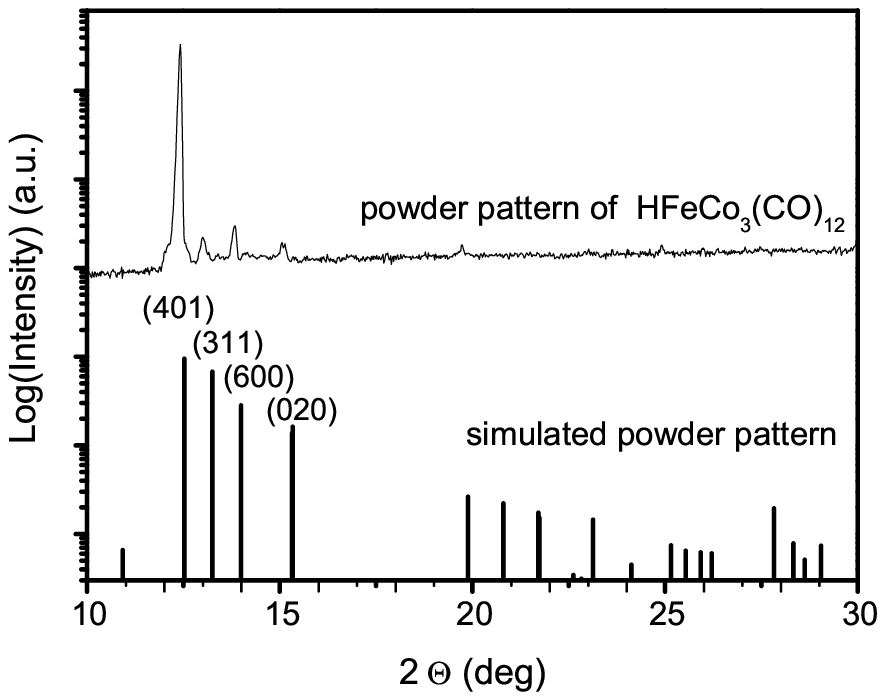}}
\caption{Simulated powder pattern and an acquired powder pattern of a re-crystallized HFeCo$_3$(CO)$_{12}$ sample containing large crystals with highly anisotropic morphology.}
\label{Powder_XRD}\end{figure}

\begin{figure}\center{\includegraphics[width=12cm]{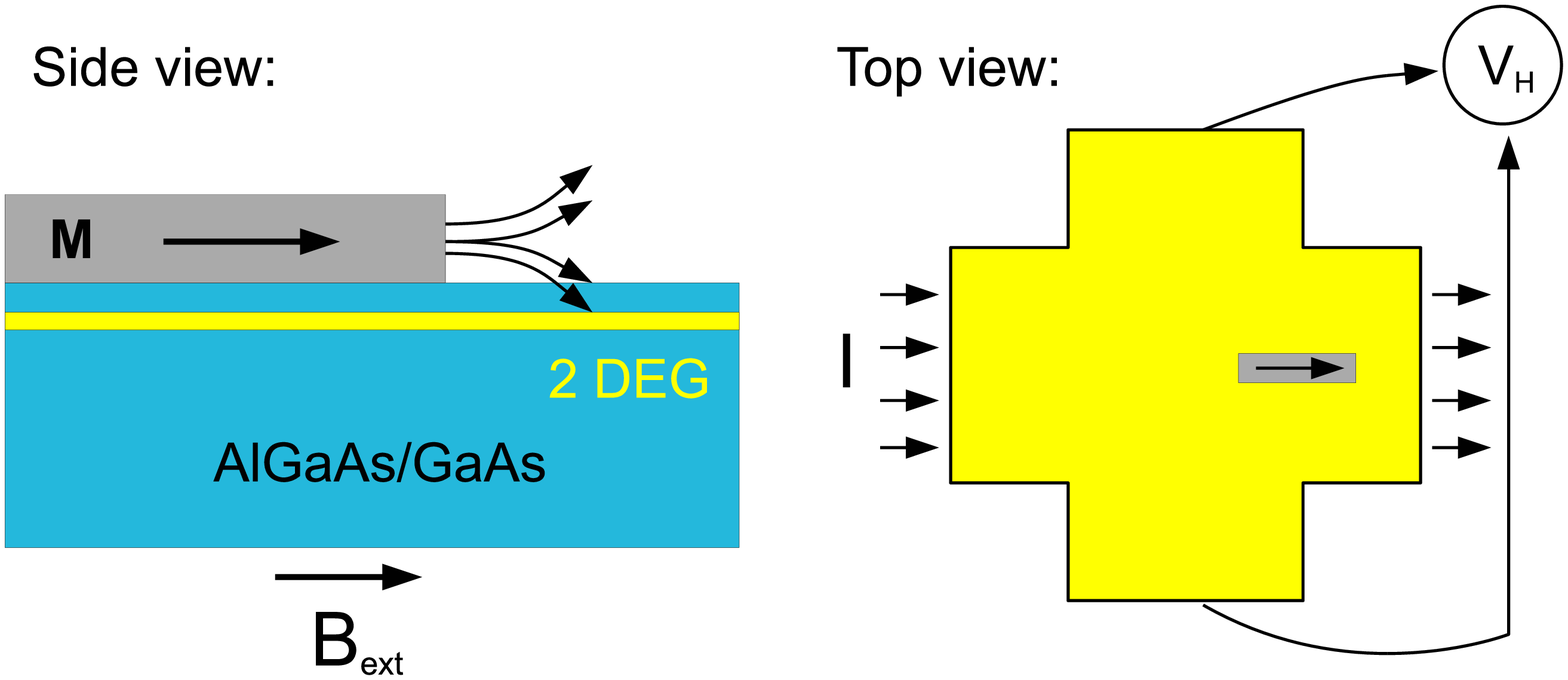}}
\caption{Sketch of the micro-Hall magnetometer used to study the magnetization reversal of the CoFe nanobar by detecting the \textit{z}-component of the stray field averaged on the 1$\times$1~$\mu m^2$ active area of the Hall cross, $<B_z>$, emanating from the sample. The micro-Hall sensor was fabricated by electron-beam lithography and wet chemical etching from a two dimensional electron system (2DES) based on an AlGaAs/GaAs heterostructure. The material combines a high electron mobility ($\mu\approx5\times10^5cm^2/Vs$) with a moderate charge carrier density ($n\approx 2.5\times10^{11}~cm^{-2}$), which is rather temperature independent between T = 4.2~K and T = 75~K, where the sensor's sensitivity is largest.}
\label{Hall_magnetometry_S3}\end{figure}

\begin{figure}[ht]\center{\includegraphics[width=12cm]{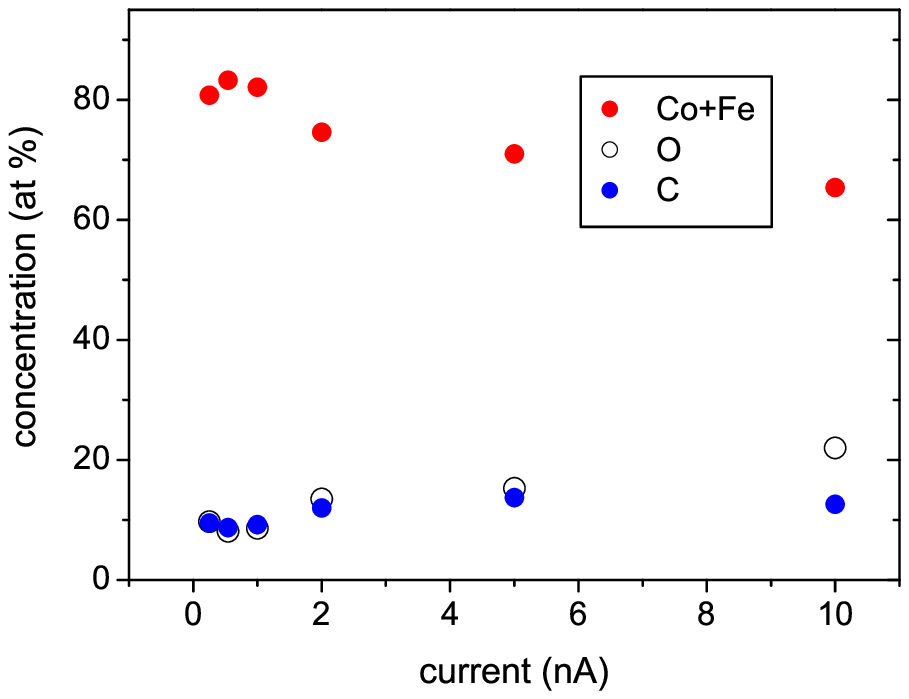}}
\caption{Atomic concentrations vs electron beam current for 5~keV electron beam voltage.}
\label{concentration1}\end{figure}

\begin{figure}[ht]\center\hspace{-15mm}{\includegraphics[width=12cm]{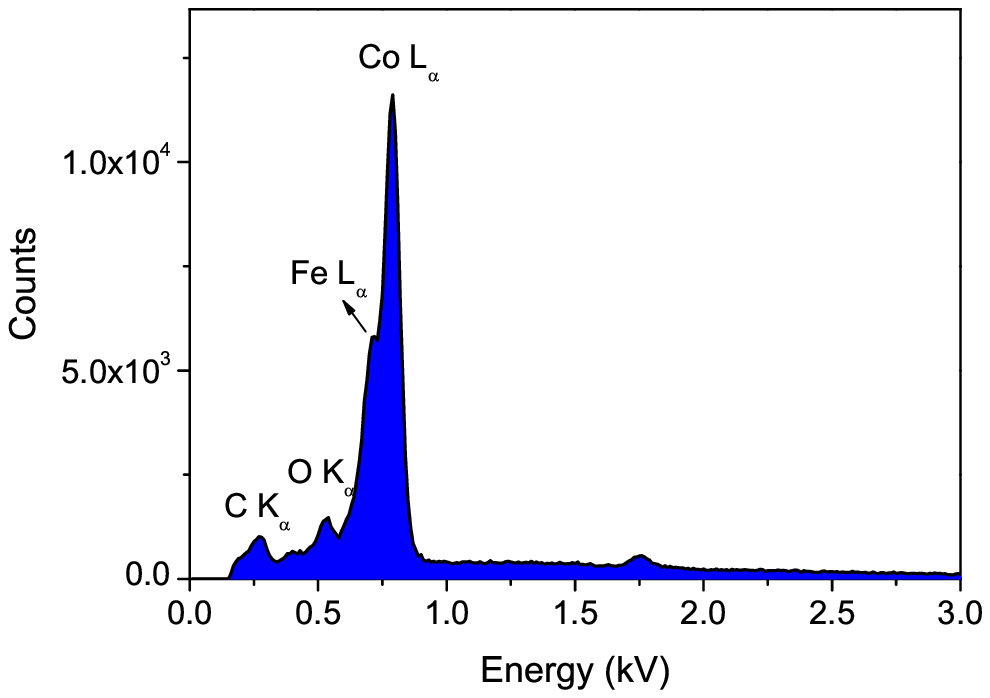}}
\caption{Typical EDX spectrum obtained by the deposition of a 2$\mu m \times$2$\mu m$ structure with electron beam parameters of 5~keV, 1~nA, 20~nm and 1~$\mu s$ for voltage, current, pitch and dwell time, respectively. The atomic element concentration is: metal 82.1$\%$, carbon 9.3$\%$, oxygen 8.6~$\%$. The measurement was carried out by using 5~keV electron beam voltage. The small Si K-line peak at 1.78~keV comes from the substrate.}
\label{edx}\end{figure}

\begin{figure}\center{\includegraphics[width=10cm]{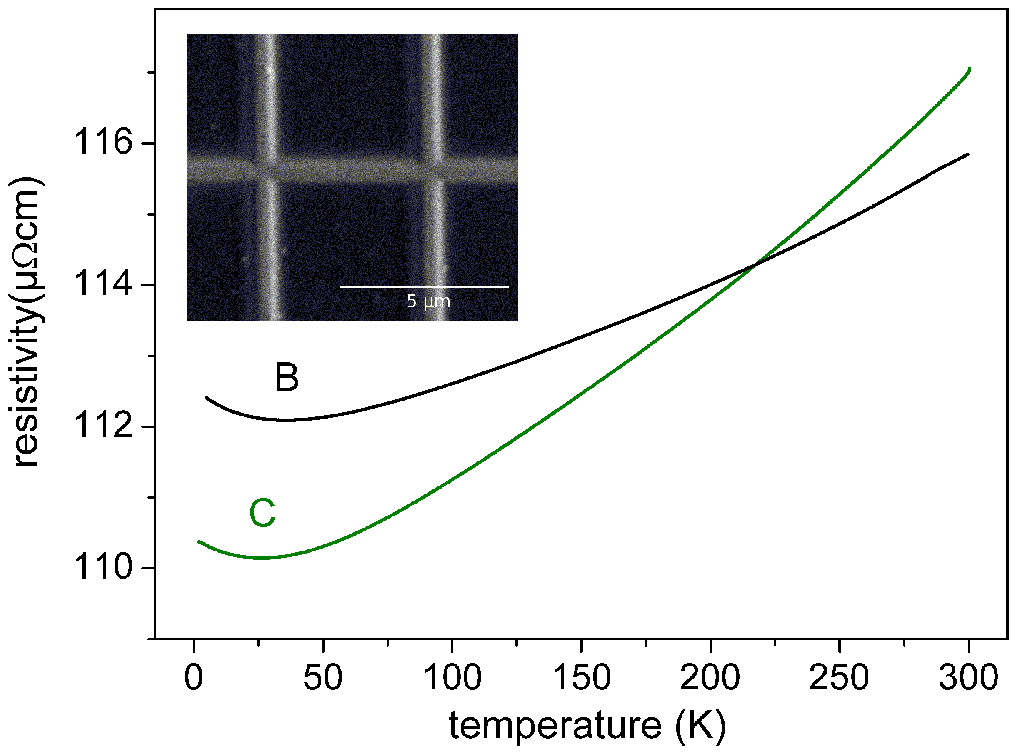}}\caption{Temperature dependence of the resistivity of two nanowires of width about 700~nm, thickness 90~nm and length 5~$\mu$m, see inset. The nanowires where prepared with the same electron beam parameters. The nanowire B was treated with the electron beam after deposition to improve the conductivity. Therefore, the resistivity at room temperature is lower for sample B than for sample C. Furthermore, since the metal content of sample C is about 2 at$\%$ higher than the one of sample B, i.e., 79~at$\%$ and 77~at$\%$, there is a difference in the RRR, i.e., 1.06 to 1.03. These values are smaller than that of sample A. Although metallic, samples B and C are not far away from the metal-insulator transition. This is also evident from the presence of the minimum at low temperature, which does not relate to a size effect as for sample A, but to a change in transport regime from metallic towards variable range hopping at low temperature. Note also that in sample C the minimum shifts to lower temperature, as expected for more metallic samples.}\label{nanowires}\end{figure}

\begin{figure}\center{\includegraphics[width=12cm]{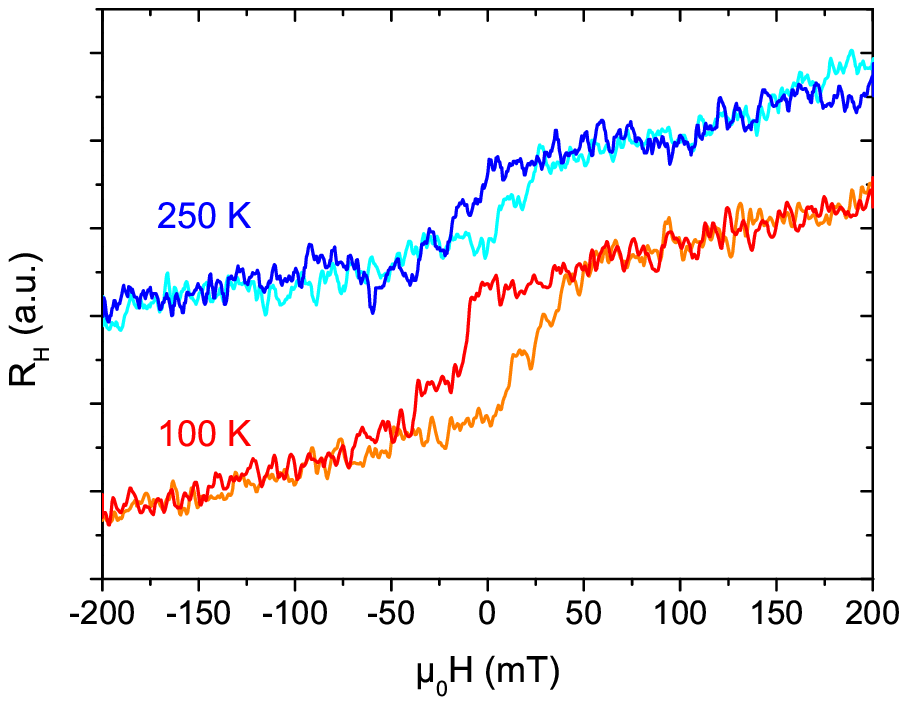}}
\caption{Hysteresis in the magnetic stray field of the CoFe alloy nanobar measured at elevated temperatures by means of a variable temperature insert (VTI) cryostat. These measurements prove the ferromagnetic nature of the material up to the highest measured temperature of 250~K. An empty reference cross was subtracted as background.}
\label{Hall_magnetometry_S7}\end{figure}

\begin{figure}\center{\includegraphics[width=14cm]{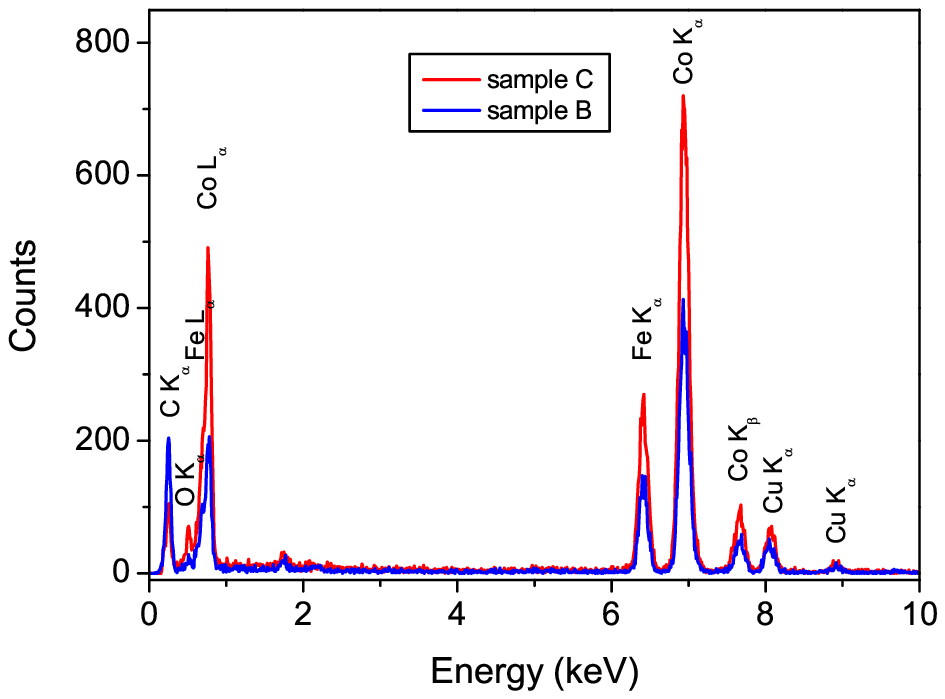}}
\caption{EDX-spectra of samples B and C obtained in the scanning transmission electron microscope (STEM) by using the high-angle annular dark-field imaging (HAADF) technique. The analysis of the spectra, which is obtained by using both the L-lines and the K-lines of Co and Fe, shows that the ratio [Co]/[Fe] is equal to three, as expected from the stoichiometry of the HFeCo$_3$(CO)$_{12}$ precursor. The small amount of the Cu signal is a result of using a copper grid for the FIB-lamella.}
\label{STEM}\end{figure}

\end{document}